\documentstyle[12pt]{article}

\newcommand{\be}{\begin{equation}}
\newcommand{\ee}{\end{equation}}
\newcommand{\bea}{\begin{eqnarray}}
\newcommand{\eea}{\end{eqnarray}}

\renewcommand{\theequation}{\arabic{section}.\arabic{equation}}

\begin{document}
\begin{titlepage}


\vspace{1in}

\begin{center}
\Large
{\bf Inflationary and Deflationary Branches 
in Extended Pre--Big Bang Cosmology}

\vspace{1in}

\normalsize

\large{James E. Lidsey$^1$}

\normalsize
\vspace{.7in}

{\em Astronomy Unit, School of Mathematical 
Sciences,  \\ 
Queen Mary \& Westfield, Mile End Road, LONDON, E1 4NS, U.K.}

\end{center}

\vspace{1in}

\baselineskip=24pt
\begin{abstract}
\noindent
The pre--big bang cosmological scenario is studied within 
the context of the Brans--Dicke theory of gravity. 
An epoch of superinflationary expansion may 
occur in the pre--big bang phase of the Universe's history 
in a certain region of parameter space. 
Two models are considered that contain a cosmological constant 
in  the gravitational and  matter sectors of the theory, respectively. 
Classical pre-- and post--big bang solutions are found for both 
models. The existence of a curvature singularity forbids a classical 
transition between the two branches. On the other hand, a quantum 
cosmological approach based on the tunneling boundary 
condition results in a non--zero transition probability. 
The transition may be interpreted as a spatial 
reflection of the wavefunction in minisuperspace.

\end{abstract}

PACS Number(s): 98.80.Bp, 98.80.Cq, 98.80.Hw 

\vspace{.7in}
$^1$Electronic mail: jel@maths.qmw.ac.uk
 
\end{titlepage}


\section{Introduction}
   
\setcounter{equation}{0}

\def\theequation{\thesection.\arabic{equation}}

Inflationary cosmology provides 
a plausible and, in principle, experimentally testable picture 
for the very early Universe \cite{inflation}. As well as resolving 
the well known problems of the hot big bang model --  
such as the horizon and flatness problems --  it results in  the 
quantum mechanical generation of primordial scalar (density) and 
tensor (gravitational wave) fluctuations \cite{scalar,tensor,SL}. 
It is widely thought that these perturbations 
may be responsible for the observed anisotropy in the 
temperature distribution of the cosmic microwave background 
radiation \cite{cmb}. Moreover, the growth of the scalar fluctuations 
via gravitational instability may have led to the formation 
of large--scale structure in the Universe. (For recent reviews see, e.g., 
Refs. \cite{review}). 

The defining feature of inflation is the acceleration 
of the cosmological scale factor, $\ddot{a} >0$. This is in contrast 
to the decelerating expansion that is characteristic of  the big bang model. 
In the standard, chaotic inflationary Universe, the 
acceleration is driven by the potential energy of a 
self--interacting scalar field \cite{linde}. The potential 
plays the role of an effective cosmological 
constant, thereby leading to a quasi--exponential expansion 
of the Universe. 

An interesting alternative to the standard inflationary Universe 
has recently been developed within the context of string theory
\cite{string}. 
It is generically 
referred to as {\em pre--big bang cosmology}. It is known that the 
tree--level, $(d+1)$--dimensional (super)string effective action 
exhibits an $O(d,d)$ symmetry \cite{dd}. Embedded in this group 
is a symmetry known as {\em scale factor duality} 
\cite{scalefactorduality}. This symmetry 
applies when the Universe is spatially flat and
homogeneous and leaves the action 
invariant under the simultaneous interchange 
$\tilde{a}= 1/a$ and $\tilde{\Phi}=  \Phi  -2 d \ln a$, where 
$\Phi$ represents the dilaton field. 

Scale factor duality therefore relates expanding cosmologies to 
contracting ones. Since the theory is also invariant under time 
reversal, $\tilde{t}=  -t$, 
the contracting branch  may itself be mapped onto a new, expanding branch.  
These two expanding solutions  are referred to as the pre-- 
and post--big bang branches, respectively. They are 
conventionally defined over the 
negative and positive halves of the time 
axis and are separated by a curvature singularity at $t=0$. 

Applying the duality transformation simultaneously 
with time reversal implies that 
the Hubble expansion parameter $H \equiv d(\ln a) /dt$ 
remains invariant, 
$\tilde{H}(-t) = H(t)$, whilst its first derivative 
changes sign, $\dot{\tilde{H}} (-t) = - \dot{H} (t)$. 
A decelerating, post--big bang solution -- characterized 
by $\dot{a}>0$, $\ddot{a}<0$ and $\dot{H}<0$ --  is therefore mapped onto 
a pre--big bang phase of inflationary expansion, since $\ddot{a} /a
= \dot{H}+H^2 >0$. The Hubble radius $H^{-1}$ 
decreases with increasing time and the expansion is therefore 
superinflationary. It  
is driven by the kinetic energy of the dilaton field. 

Since the two branches are separated by a curvature singularity, however, it 
is not clear how the 
transition between the pre-- and post--big bang phases 
might proceed. This is the graceful exit problem of 
the pre--big bang scenario. Such a problem is characteristic 
of more general inflationary scenarios that are driven 
by the dilaton's kinetic energy \cite{levinnogo}. 
One possible solution 
is to introduce a suitable dilaton potential that modifies the 
classical solution around $t \approx 0$ \cite{potential}. 
It has recently been shown, however, 
that a branch change can not occur for a realistic dilaton potential
if one is limited to the lowest--order expansion of the 
string action \cite{modify}. 
An alternative approach would be to employ the techniques 
of conformal field theory, where all higher--order terms in the action 
are considered \cite{KK}. Unfortunately, however, 
the appropriate conformal background is presently unknown. 

In view of this, it has recently been proposed that the graceful 
exit problem might be addressed by employing a quantum cosmological 
approach \cite{GMV,GV}. It is reasonable to suppose that 
quantum gravitational effects should become 
important in the high curvature regime and, indeed,  
the relationship between 
quantum cosmology and string theory has been considered 
previously by a number of authors
Ref. \cite{sqc}. The significance  of scale factor 
duality in this approach has also been discussed 
\cite{BB,L} and ann 
$O(d,d)$--invariant Wheeler--DeWitt 
equation was recently 
derived by Gasperini, Maharana and Veneziano \cite{GMV} and by
Kehagias and Lukas \cite{KL}. 

Gasperini {\em et al.} considered the minisuperspace model 
for the spatially, homogeneous 
Bianchi I Universe \cite{GMV}. 
They found that the wavefunction of the Universe could be 
expanded in terms of plane waves in minisuperspace, where  
the configurations associated with the pre-- and post--big bang branches 
corresponded to waves moving in {\em opposite} directions along 
the effective spatial coordinates. This is important, because it suggests 
that a transition between the pre--and post--big bang phases 
might be possible if the wavefunction undergoes a spatial reflection in 
minisuperspace. 

Motivated by these considerations, we investigate 
whether the concept of pre--big bang cosmology can be 
extended beyond the truncated string effective action
to include more general dilaton--graviton systems. 
Theories of this type are interesting in their own right and  they 
also place the results and predictions of string cosmology in a wider setting.
We shall consider the vacuum Brans--Dicke theory 
of gravity \cite{BD}, whose action is given by
\be
\label{action}
S=\int d^4x \sqrt{-g} e^{-\Phi} \left[ R - \omega \left( \nabla \Phi 
\right)^2 -2 \Lambda (\Phi) \right]
\ee
where the metric 
$g_{\mu\nu}$ has  signature $(-,+,+,+)$, $R$ is the Ricci curvature scalar 
and $g \equiv {\rm det} g_{\mu\nu}$. 
The parameter $\omega$
determines the strength of the coupling between the 
dilaton and graviton degrees of freedom and  is assumed to be 
a space--time constant. 
The function $\Lambda(\Phi )$ determines the self--interactions of the 
dilaton. The truncated string effective action is 
given by Eq. (\ref{action}) with $\omega =-1$ and constant $\Lambda$. 

We will assume that $g_{\mu\nu}$ represents the space--time of the physical 
Universe and will therefore work in the Jordan frame rather than 
the Einstein frame. We will consider the
spatially flat, isotropic and homogeneous cosmology  
with a line element $ds^2 =-N^2(t)dt^2 +e^{2 \alpha (t)} d{\rm {\bf x}}^2$, 
where $e^{\alpha (t)}$ represents the scale factor of the Universe and 
is a function of cosmic time $t$ only and 
$N(t)$ is the lapse function. We further assume that the dilaton field 
is constant on the surfaces of homogeneity and 
consider the region of parameter space where $\omega > -3/2$ and 
$\Lambda (\Phi) \ge 0$. Thus, the weak energy condition is always satisifed. 

We begin in Section 2 by considering the special case where the 
dilaton potential vanishes, $\Lambda =0$. 
Such a model provides a useful framework within which the key ideas of duality 
and branch changing can be discussed. We identify 
the region of parameter space that leads to a 
superinflationary, pre--big bang phase. We then 
employ a generalized scale factor duality 
to illustrate how the pre-- and post--big bang branches are 
related at both the classical and quantum levels. In particular, it is shown 
how the two branches correspond to left-- and right--moving  waves in 
minisuperspace. A reflection of the wavefunction is only 
possible, however, if a dilaton potential is included. We therefore 
introduce a cosmological constant into the model in Section 3 by specifying 
$\Lambda={\rm constant}$. This model is quantized in Section 4 and 
the conditional probability for a reflection of the wavefunction is 
calculated. We then proceed in Section 5 to consider a 
second model  where $\omega =-1/2$ and $\Lambda \propto e^{\Phi}$. 
This potential plays the role of 
a cosmological constant in the matter sector of the theory. 
It is found that the probability for a transition between the pre-- and 
post--big bang branches is also non--zero. We conclude 
in Section 6. 

 Units are chosen such that $\hbar = c =1$.

\section{Scale Factor Duality and 
 Pre--Big Bang Cosmology in Brans--Dicke Theory}

\setcounter{equation}{0}

\def\theequation{\thesection.\arabic{equation}}

Eq. (\ref{action}) simplifies to 
\be
\label{simpleaction}
S=\int dt e^{3\alpha -\Phi} \left[ \frac{1}{N} \left( -6 \dot{\alpha}^2 
+6 \dot{\alpha}\dot{\Phi} +\omega \dot{\Phi}^2 \right) 
-2 N \Lambda (\Phi) \right]
\ee
after integration over the 
spatial variables, where it has been assumed that the spatial sections 
have finite volume, 
a boundary term has been neglected and the dilaton has been rescaled 
$\Phi \rightarrow \Phi -\ln \int d^3x$. 
It has recently been shown that kinetic sector of this action  
is invariant under a generalized scale factor duality
\bea
\label{sfd}
{\alpha} = \left( \frac{2+3\omega}{4+3\omega} \right) \tilde{\alpha} 
-\left( \frac{2(1+\omega ) }{4+3\omega} \right) 
\tilde{\Phi} \nonumber \\
{\Phi} =- \left( \frac{6}{4+3\omega} \right) \tilde{\alpha} 
-\left( \frac{2+3\omega}{4+3\omega} \right) \tilde{\Phi}
\eea
when $\omega \ne -4/3$ \cite{L}. This descrete symmetry is embedded 
within a more general,   continuous Noether symmetry that exists  in the 
cosmological field equations of 
Brans--Dicke theory \cite{L1996}. It reduces to the scale factor duality 
associated with the truncated string effective action when $\omega =-1$. 

The free field model $(\Lambda =0)$ represents the limiting case
of a more general class of models in which the dilaton potential 
becomes negligible near the curvature singularity. It is also 
relevant to the recently studied kinetic inflationary scenario, 
where the acceleration of the scale factor is driven by the kinetic 
energy of the dilaton field rather than its potential energy 
\cite{levinnogo,kinetic,kinetic1}. 
The classical field equations derived from action 
(\ref{simpleaction}) for $\Lambda =0$ are given by 
\bea
\label{0field1}
\ddot{\Phi} -\dot{\Phi}^2 +3\dot{\alpha}\dot{\Phi} = 0 \\
\label{1field2}
\dot{\alpha}^2 -\dot{\alpha}\dot{\Phi} -\frac{\omega}{6} 
\dot{\Phi}^2  = 0
\eea
for $N=1$. They admit the  power--law solution 
\be
\label{vacuum}
e^{\alpha} = e^{\alpha_0}
|t|^{p_{\pm}}, \qquad e^{\Phi} = e^{\Phi_0} |t|^{3p_{\pm} -1}
\ee
where 
\be
\label{p+-}
p_{\pm} \equiv \frac{1}{4+3\omega} \left[ 1+\omega \pm \left( 
1+\frac{2\omega}{3} \right)^{1/2} \right]
\ee
The integration constants 
have been chosen so that the curvature singularity is at $t=0$. 
There are two distinct solutions 
depending on whether $p_{\pm}$ is chosen. It 
may be verified by direct substitution that the two branches are 
related by the duality transformation  (\ref{sfd}) and the duality 
is effectively generated by the simultaneous interchange 
$p_+ \leftrightarrow p_-$. 

The square root of the Friedmann equation (\ref{1field2}) implies 
that the Hubble expansion parameter is given by 
$H=\dot{\Phi} (1\pm f)/2$, where $f \equiv [1+2\omega/3]^{1/2}$. It
can then be shown that the scale factor accelerates if and only if 
$f \pm 1< 0$ \cite{kinetic1}.  Thus, kinetic inflation can only occur if the 
negative root is chosen and the coupling constant is negative, $\omega <0$.  
One may further show that 
\be
\label{HDOT} 
\dot{H} = - \frac{1}{4} \dot{\Phi}^2 (1\pm 3f) (1\pm f)
\ee
thereby implying that superinflation 
with $\dot{H}>0$ is possible when $-4/3 < \omega <0$. 
It is this region of parameter space that is of relevance to the 
pre--big bang scenario. The time--reversed negative root $(t <0, p =
p_-)$ corresponds to a superinflationary,  pre--big bang branch. 
It is equivalent to the pole--law inflation considered by 
Pollock and Sahdev \cite{PS}. The 
positive  root $(p = p_+)$ then represents the deflationary, 
post--big bang phase. The two branches are shown in Figure 1. 
We shall refer to them as the $(+)$-- and $(-)$--branches, respectively. 

\vspace{.1in}

\centerline{\bf Figure 1}

\vspace{.1in}

It proves convenient to define the new coordinate pair
\bea
\label{newcoords}
\beta \equiv \sqrt{\frac{6}{4+3\omega}} \left[ \alpha +(1+\omega ) \Phi 
\right] \nonumber \\
\sigma \equiv \kappa^{-1} (\Phi -3 \alpha ) , \qquad \kappa \equiv 
\sqrt{\frac{4+3\omega}{6+4\omega}}
\eea
It follows that the scale factor duality 
transformation (\ref{sfd}) is generated by the simultaneous 
interchange $\tilde{\sigma} =\sigma$ and $\tilde{\beta} =-\beta$. 
The pre-- and post--big bang classical trajectories (\ref{vacuum}) are 
then given by $\beta =\beta_0 \pm \kappa^{-1} \ln (\pm t)$ and $\sigma 
= \sigma_0 - \kappa^{-1} \ln (\pm t)$, respectively, where $\{ \beta_0 , 
\sigma_0 \}$ are related to $\{ \alpha_0 ,\Phi_0 \}$. 
Action (\ref{simpleaction}) 
simplifies to $S=\frac{1}{2} \int dt e^{-\kappa \sigma} [ 
\dot{\beta}^2 - \dot{\sigma}^2 ]$ and  
the momenta conjugate to $\beta$ and $\sigma$ are given by
$p_{\beta} =\dot{\beta}e^{-\kappa \sigma}$ and $p_{\sigma} =- 
\dot{\sigma}e^{-\kappa\sigma}$, respectively. 
We may conclude, therefore, that since 
$\sigma$ is invariant under 
scale factor duality, the pre-- and post--big bang branches 
have equal and {\em opposite} momentum along the 
$\sigma$--axis, i.e., $p_{\sigma}^{(+)} = - p_{\sigma}^{(-)}
= - 1/\kappa$. (We specify $\sigma_0 =0$ for simplicity). On the 
other hand, the momentum 
conjugate to $\beta$ is the same for both branches, 
$p^{(+)}_{\beta} 
= p^{(-)}_{\beta} = 1/\kappa$. 

It is this feature that allows us to interpret the two branches at the 
quantum level in terms of left-- and right--moving waves. 
The Wheeler-DeWitt equation is the operator version of the 
Hamiltonian constraint $NH_0= p_{\beta}\dot{\beta} + p_{\sigma}\dot{\sigma}
-L= 0$, where $L$ 
is the Lagrangian density \cite{WDWequation}. 
It is derived in the usual fashion by 
identifying the conjugate momenta $p^{\mu}$ with the operators $p^{\mu} = 
\pm  i \partial /\partial q_{\mu}$. 
The wavefunction of the Universe is an eigenvector
of the Hamiltonian operator and physical states have zero eigenvalue, i.e., 
$H_0 \Psi =0$. If one ignores ambiguities in the operator ordering, this 
constraint takes the form of the one--dimensional wave equation 
\be
\label{wheelerdewitt}
\left[ \frac{\partial^2}{\partial \sigma^2} - 
\frac{\partial^2}{\partial \beta^2} \right] \Psi  = 0
\ee
We may therefore view the wavefunction as a free, bosonic particle 
propagating over $(1+1)$--dimensional Minkowski space--time, 
where $\beta$ and $\sigma$ represent the time--like and space--like 
variables, respectively. 

The general solution to this equation has two components 
$\Psi =\Psi^{(+)}(u_+)+\Psi^{(-)}(u_-)$, where 
$\Psi^{(\pm )}$ are arbitrary functions of $u_{\pm} 
\equiv  \beta \mp \sigma$. We may consider wavefunctions of the form 
$\Psi^{(\pm )} =\exp [i S_{\pm} ]$, where $S_{\pm} =S_{\pm} (u_{\pm}) $ 
are solutions to the 
Hamilton--Jacobi equation $(\partial_{\sigma}S_{\pm})^2 = 
(\partial_{\beta}S_{\pm})^2$. 
These wavefunctions are peaked around 
the classical trajectories given by $p_{\beta}^{(\pm)} = \partial 
S_{\pm} /\partial \beta$ and $p_{\sigma}^{(\pm)} =\partial S_{\pm} /
\partial {\sigma}$. Specifying $S_{\pm} = \kappa^{-1} u_{\pm}$ therefore 
implies that $\Psi^{(+)}$ and $\Psi^{(-)}$ are the wavefunctions 
for the pre-- and post--big bang branches, respectively. Moreover, 
they may be viewed as plane waves moving in opposite spatial directions 
through minisuperspace. This suggests that a spatial reflection 
of the wavefunction in minisuperspace could 
result in a transition between the pre--and post--big bang branches. 
Such a reflection can not proceed for this model, however, 
because the dilaton field is free. The simplest extension 
is to allow this field to self--interact.
This interaction  will manifest itself as an effective potential in 
the Wheeler--DeWitt equation and it is this potential that may lead to 
a reflection of the wavefunction. 

\section{A Cosmological Constant in the Gravitational Sector}

\setcounter{equation}{0}

\def\theequation{\thesection.\arabic{equation}}

In this Section we introduce a cosmological constant into the gravitational 
sector of the theory by  assuming that $\Lambda (\Phi) \equiv \Lambda$
is a constant in Eq. (\ref{simpleaction}). 
The generalized scale factor duality (\ref{sfd}) 
is respected by this model since the potential 
energy in the Lagrangian has the form $e^{-\kappa\sigma}$.

The field equations derived from action (\ref{simpleaction}) are 
\bea
\label{field1}
\ddot{\alpha}+3\dot{\alpha}^2 -\dot{\alpha}\dot{\Phi} 
= \frac{2 (1+\omega )\Lambda}{2\omega +3} \\
\label{field2}
\ddot{\Phi} -\dot{\Phi}^2 +3\dot{\alpha}\dot{\Phi} = - 
\frac{2\Lambda}{2\omega +3} \\
\label{field3}
\dot{\alpha}^2 -\dot{\alpha}\dot{\Phi} -\frac{\omega}{6} 
\dot{\Phi}^2  =
\frac{\Lambda}{3}
\eea
Now, multiplying Eq. (\ref{field1}) by $3$ and substracting 
Eq. (\ref{field2}) 
implies that 
\be
\label{field4}
\frac{d^2}{dt^2} \left( e^{-\varphi} \right) = \eta^2 e^{-\varphi} 
\ee
where 
\be
\eta \equiv  \left[ 2\Lambda \left( \frac{4+3\omega}{2\omega+3} \right) 
\right]^{1/2}
\ee
and $\varphi \equiv \kappa \sigma$ 
represents a shifted dilaton field \cite{UK,CR}. 
The general solution to Eq. (\ref{field4}) 
is given by
\be
\label{sol1}
e^{-\varphi} = Ae^{\eta t}+Be^{-\eta t}
\ee
where $A$ and $B$ are arbitrary constants. 

We may rewrite Eq. (\ref{field2}) in terms of the shifted dialton: 
\be
\label{field5}
\frac{d}{dt} \left( \frac{d \Phi}{dt} e^{-\varphi} \right) = - 
\frac{2\Lambda}{2\omega +3} e^{-\varphi}
\ee
This equation admits the first integral 
\be
\label{firstintegral}
\dot{\Phi} = \frac{1}{4+3\omega} \dot{\varphi} + c e^{\varphi}
\ee
where $c$ is a constant 
and substituting this result  into the Friedmann 
equation (\ref{field3}) implies that 
\be
\label{c} 
c^2 = - \frac{24 AB}{4+3\omega} \Lambda
\ee

We will consider the specific solution $A=-B$, since this implies that 
the singularity may be located at $t=0$ without 
loss of generality. Eq. (\ref{sol1}) then  implies that 
\be
\label{sol}
\varphi = \varphi_0 - \ln {\rm sinh} \left| \eta t \right|
\ee 
where $e^{-\varphi_0} \equiv 2A$. 
The functional forms of $\alpha (t) $ and $\Phi (t)$ may now 
be determined by substituting this result into Eq. 
(\ref{firstintegral}) and integrating. The 
solution that corresponds to an expanding  Universe with a vanishing 
scale factor in the limit $t \rightarrow 0^+$ is given by
\bea
\label{sol2}
e^{\alpha} =e^{\alpha_0} \left[ {\rm sinh} (\eta t/2) \right]^{p_+} 
\left[ {\rm cosh} (\eta t/2 ) \right]^{p_-} \nonumber \\
e^{\Phi} =e^{\Phi_0} \left[ {\rm sinh} (\eta t/2 ) \right]^{3p_+-1} 
\left[ {\rm cosh} (\eta t/2) \right]^{3p_--1}
\eea
where $\{ \alpha_0 ,\Phi_0 \}$ are integration 
constants. This solution is defined over the positive half of the time 
axis and there is a singularity 
in the curvature at $t=0$. 

We may now apply the scale factor duality to solution (\ref{sol2}) 
to generate the second branch. It follows that 
\bea
\label{sol3}
e^{\tilde{\alpha}} =e^{\tilde{\alpha}_0} \left[ 
{\rm sinh} (-\eta t/2) \right]^{p_-} \left[ {\rm cosh} (-\eta t/2) 
\right]^{p_+} \nonumber \\
e^{\tilde{\Phi}}=e^{\tilde{\Phi}_0} \left[ {\rm sinh} (-\eta t/2) 
\right]^{3p_- -1} \left[ {\rm cosh} (-\eta t/2) \right]^{3p_+ -1} 
\eea
where we have also performed the time reversal. This solution 
is related to Eq. (\ref{sol2}) 
by the simultaneous interchange $p_+ \leftrightarrow 
p_-$. This corresponds, in effect, to choosing the 
negative square root of Eq. (\ref{c}).
The cosmology given by (\ref{sol3}) approaches a singularity 
as $t \rightarrow 0^-$. There is also a 
singularity in the effective gravitational coupling $G_{\rm eff} 
\equiv e^{\tilde{\Phi}}$ in this limit. The solutions 
(\ref{sol2}) and (\ref{sol3}) may 
therefore be viewed as two distinct branches corresponding to 
the post-- and pre--big bang phases of the Universe's history, respectively. 

The qualitative behaviour of solution (\ref{sol2}) and its dual (\ref{sol3}) 
is shown in Figures 2a--2c for $-4/3 < \omega < 0$. 
The behaviour in the limit  $|t| \rightarrow \infty$ is given by 
\bea
\label{attractor}
\alpha_{\infty} \propto \left( \frac{2\Lambda}{(4+3\omega )(2\omega +3)}
\right)^{1/2} (1+\omega)  |t| \nonumber \\
\Phi_{\infty} \propto -\left( \frac{2\Lambda}{(4+3\omega )(3+2\omega )} 
\right)^{1/2} |t|
\eea
This represents the analogue of the 
de Sitter solution. Near the singularity, on the other hand, 
the solutions simplify to the 
free--field solution (\ref{vacuum}), since the dilaton's kinetic 
energy dominates the energy density of the Universe in this regime. For 
$-1 < \omega <0$, the post--big bang branch is initially 
deflationary, but inflates at later times when the cosmological constant 
begins to dominate the dynamics. The 
corresponding pre--big bang branch is a bouncing solution. It collapses 
from infinity, reaches a minimum size and then re--expands 
to infinity as $t \rightarrow 0^-$. When $\omega =-1$, both 
branches expand monotonically, as shown in Figure 2b. 
For $-4/3 < \omega < -1$, the pre--big bang branch expands monotonically 
to infinity, 
while its dual expands initially, but recollapses at later times. 

The behaviour of the effective gravitational coupling $e^{\Phi}$ 
is shown in Figure 3 for both branches. The coupling 
vanishes initially in  the pre--big bang branch and 
becomes infinitely strong as $t \rightarrow 0^-$. 
In the post--big bang solution, the coupling also vanishes  
initially, but falls back to zero after a certain time has elapsed. 

\vspace{.1in}

\centerline{\bf Figures 2 \& 3}

\vspace{.1in}

In the following Section, we will argue that a transition between 
the pre-- and post--big bang branches may proceed at the 
quantum level even though such a process is classically forbidden. 

\section{Quantum Transitions}
 
\setcounter{equation}{0}

\def\theequation{\thesection.\arabic{equation}}

The cosmology may be quantized by rewriting the system 
in terms of the variables (\ref{newcoords}). 
Action (\ref{simpleaction}) then takes the canonical form
\be
\label{newaction}
S=\int dt e^{-\kappa \sigma} \left( \frac{1}{2} \dot{\beta}^2 
-\frac{1}{2} \dot{\sigma}^2 -2\Lambda \right)
\ee
and the momenta conjugate to $\beta$ and $\sigma$ are given by
\bea
\label{betamomentum}
p_{\beta} =\dot{\beta}e^{-\kappa \sigma} \equiv k  \\
\label{sigmamomentum}
p_{\sigma} =- \dot{\sigma}e^{-\kappa\sigma}
\eea
respectively. Substitution of Eqs. (\ref{sol2}) and (\ref{sol3}) into 
Eq. (\ref{betamomentum}) implies that 
$k = 2 \sqrt{\Lambda} e^{-\varphi_0} = (\eta /\kappa ) e^{-\varphi_0}$. 

Since the scale factor duality (\ref{sfd}) is respected 
by action (\ref{newaction}), $p_{\beta}$ is invariant 
under duality and time reversal, 
whilst $p_{\sigma}$ changes sign. Indeed, we find that 
the momenta for the pre-- and post big bang branches are given by 
$p^{(+)}_{\beta} = p^{(-)}_{\beta} = k$ and 
\be
p^{(\pm )}_{\sigma} = \mp k  {\rm cosh} \left( \mp \eta t \right)
\ee
It follows that $p_{\sigma}^{( \pm )}$ 
asymptotically approaches a constant value
near to the singularity $( |t| \rightarrow 0)$: 
\be
\label{limitstr}
\lim_{\sigma \rightarrow +\infty} p^{( \pm )}_{\sigma} = \mp k
\ee
whereas we find that
\be
\label{limitlow}
\lim_{\sigma \rightarrow -\infty} 
p^{(\pm )}_{\sigma} = \mp ke^{\varphi_0 -\kappa \sigma}
\ee
in the low energy regime ($|t| \rightarrow \infty$, $\sigma \rightarrow 
-\infty$).

The system is quantized by identifying the conjugate momenta with 
the operators $\hat{p}_{\beta} \equiv i\partial /\partial \beta$ and 
$\hat{p}_{\sigma} \equiv i \partial /\partial \sigma$, 
respectively\footnote{The unconventional sign choice of Gasperini {\em 
et al.} is chosen in the differential operators \cite{GMV}. This ensures  that 
the positive momentum waves move from negative to positive 
values of the effective spatial variable in minisuperspace \cite{G}.}.
The question of 
operator  ordering now arises. Since $\beta$ is related 
to the logarithm of the 
scale factor, we may consider the semi--general 
operator ordering proposed by Hartle and Hawking \cite{HH}:
\be
\hat{p}_{\beta}^2 = - e^{p \beta} \frac{\partial}{\partial \beta} 
e^{-p \beta} \frac{\partial}{\partial \beta}
\ee
where $p$ is an arbitrary constant. Our choice of $p$ is motivated by 
the scale factor duality (\ref{sfd}) 
associated with the kinetic sector of the classical action 
(\ref{simpleaction}). 
It is reasonable to suppose that the Wheeler-DeWitt equation 
should respect this symmetry and should therefore 
be invariant under the interchange $\tilde{\beta} =- \beta$. In view of this, 
we should specify $p=0$ and 
this implies that the Wheeler-DeWitt equation is then given by 
\be
\label{WDW}
\left[ \frac{\partial^2}{\partial \sigma^2} -\frac{\partial^2}{\partial 
\beta^2} +4\Lambda e^{-2\kappa \sigma} \right] \Psi =0
\ee
Furthermore, Eq. (\ref{betamomentum}) implies that $\beta$ is a monotonically 
increasing function. It may therefore be viewed as the effective 
time coordinate of the minisuperspace. The space--like variable may 
then be identified with $\sigma$. 

The wavefunction is an eigenstate of the momentum operator $\hat{p}_{\beta}$ 
and we specify the eigenvalue to be $k$, in agreement 
with Eq. (\ref{betamomentum}). 
The general, separable solution to Eq. (\ref{WDW}) is 
then given by 
\be
\label{WDWsol1} 
\Psi = Z_{\pm i k/\kappa } (z) e^{ -ik\beta}
\ee
where $z \equiv (2 \sqrt{\Lambda}/\kappa) e^{-\kappa \sigma}$ and $Z_{\pm
ik/\kappa}$ is a linear combination of Bessel functions 
of order $\pm ik/\kappa$. 

The specific solution appropriate to the pre--big bang scenario 
is determined from the tunneling boundary 
condition \cite{GMV,tunnel,vilenkin}. Vilenkin has formulated this condition 
as a boundary condition on the superspace 
of all three--metrics and matter configurations \cite{vilenkin}. 
In this formulation,  the wavefunction of the Universe 
should be everywhere bounded and consist only of outgoing modes 
at the singular boundaries of superspace that correspond  to singularities 
in the four--geometry. 
The minisuperspace considered in this model is 
$(1+1)$--dimensional Minkowski space--time and  
the outward (inward) trajectories  at the singular 
boundary of this minisuperspace correspond to the pre-- (post--) big bang 
branches, respectively. 
The tunneling boundary condition is therefore consistent with the pre--big 
bang initial conditions. 

In the vicinity of the 
singularity $(\sigma \rightarrow +\infty)$, 
the effective potential in the Wheeler--DeWitt 
equation (\ref{WDW}) becomes negligible and this equation therefore 
reduces to the free wave equation (\ref{wheelerdewitt}). 
Consequently, the general solution of Eq. (\ref{WDW}) 
in this limit
may be expanded in terms of plane waves, i.e., 
$\Psi^{(\pm )}_{+\infty} \propto \exp [-ik (\beta \mp \sigma )]$. 
It follows that 
the wavefunction is also an eigenvector of the momentum operator 
$\hat{p}_{\sigma}$, i.e., 
$\lim_{\sigma \rightarrow +\infty} \hat{p}_{\sigma} \Psi^{(\pm )}_{+\infty} = 
\mp k \Psi^{(\pm )}_{+\infty}$. We may conclude 
from Eq. (\ref{limitstr}), therefore, that 
$\Psi_{+\infty}^{(+)}$ and $\Psi_{+\infty}^{(-)}$ represent wavefunctions 
for the pre-- and post--big bang branches in the high  energy limit. 

The wavefunction must therefore reduce to  $\Psi = \Psi^{(+)}_{+\infty}$ 
at the singular boundary of minisuperspace 
in order to satisfy the tunneling boundary condition. 
It is this condition that then determines the order of the Bessel 
function in Eq.  (\ref{WDWsol1}). In the limit 
that $z \rightarrow 0$, 
$J_p (z) \propto z^p$,  and this implies that we should choose 
$p= -ik/\kappa$. The solution 
to Eq. (\ref{WDW}) that represents the quantum 
version of the pre--big bang branch on the approach to 
the curvature singularity is therefore given by
\be
\label{WDWsol2} 
\Psi = J_{-ik/\kappa} \left( 2\sqrt{\Lambda}/\kappa e^{-\kappa\sigma}
\right) e^{ -ik\beta}
\ee
modulo a constant of proportionality. 

We may now discuss the low energy limit $(z \rightarrow +\infty)$. 
The form of the Bessel function in this limit implies that the wavefunction 
may be expanded as a linear 
superposition  of left-- and right--moving modes of the form \cite{AS}
\be
\lim_{\sigma \rightarrow -\infty} \Psi = \Psi^{(+)}_{-\infty} + 
\Psi^{(-)}_{-\infty}
\ee
where
\be
\label{WDWsol3}
\Psi_{-\infty}^{(\pm )} \equiv  \left( 2\pi z \right)^{-1/2} 
\exp \left[ -ik \beta \mp i z \pm i \frac{\pi}{4}
\pm \frac{\pi k}{2\kappa} \right]
\ee
The wavefunctions 
(\ref{WDWsol3}) are eigenstates of the momentum operator $\hat{p}_{\sigma}$, 
since the prefactor is a slowly varying function. We find that  
\be
\lim_{\sigma \rightarrow -\infty} \hat{p}_{\sigma} 
\Psi^{(\pm )}_{-\infty} = 
\mp k e^{\varphi_0 -\kappa \sigma} \Psi^{(\pm )}_{-\infty}
\ee
and comparison with Eq. (\ref{limitlow}) 
then implies that 
$\Psi^{(\pm )}_{-\infty}$  represent the wavefunctions for the pre-- 
and post--big bang phases in the low energy, weak 
coupling  limit. 

The wavefunction in this regime is therefore a 
linear superposition of the pre-- and post--big bang components.
Since these components are moving in opposite spatial directions through 
minisuperspace, we may consider the probability for 
the wavefunction to undergo a spatial reflection \cite{GMV}. This is 
given by the reflection coefficient and is defined as
the ratio of the 
current density of the reflected modes to the current 
density of the incident modes:
\be
R \equiv  \frac{\left| \Psi^{(- )}_{-\infty} \right|^2}{\left| 
\Psi^{(+)}_{-\infty} \right|^2}
\ee
This ratio represents the probability for a branch 
change to occur between the pre-- and post--big bang phases and 
for the model considered in this Section, we find that 
\be
\label{R}
R=e^{-2 \pi k/\kappa}
\ee

The approximate dependence of the transition 
probability on the cosmological constant may be estimated 
from the definition of $k$ given by Eq. (\ref{betamomentum}). We 
evaluate this expression at a time $t_s <0$ defined by the condition 
$H(-t_s) = {\cal{O}} (1)$ \cite{GMV}. 
We will assume that $| \eta t_s | \ll 1$ and 
this is valid if $\Lambda$ is not too large. It follows from 
Eqs. (\ref{betamomentum}) and (\ref{newcoords}) that 
\be
\label{k}
k = \frac{\dot{\beta}_s \Omega (\alpha_s)}{g_s^2}
\ee
where $g_s^2 \equiv e^{\Phi_s}$ 
is the effective coupling at the scale $t=t_s$ and 
$\Omega  (\alpha_s )$ is the proper spatial volume at that time. The 
approximate form for $\beta$ is evaluated from Eqs. (\ref{newcoords}) 
and (\ref{sol3}): 
\be
\label{betas}
\beta_s \approx - \kappa^{-1} \ln \left( -\eta t_s /2 \right)
\ee
and the direct dependence on $t_s$ is eliminated by 
evaluating $H (-t_s)$. The proper volume $\Omega_s$ is calculated 
by normalizing to the initial proper volume $\Omega_i$ at 
$t=-\infty$ and employing Eq. (\ref{sol3}). Substituting the result into 
Eq. (\ref{k}) then implies that 
\be
R \approx \exp \left[ \frac{2\pi}{p_-} \frac{\Omega_i}{g_s^2} 
\left( \frac{4 \omega +6}{4 +3 \omega} \right) \left( - \frac{\eta p_-}{2} 
\right)^{3p_-} \right]
\ee
We recall that $p_- <0$ when $-4/3 < \omega <0$. The dependence 
of the probability on $\Lambda$ is therefore 
\be
\label{approximate}
R \approx \exp \left[ - \frac{1}{\Lambda^{3|p_-|/2}} \right]
\ee
and this suggests that larger values of the cosmological constant are 
favoured. We should emphasize, however, 
that this expression is only valid for small $\Lambda$ and a 
more accurate calculation would be required to determine the 
general dependence. When $\omega =-1$, Eq. (\ref{approximate}) 
exhibits the same $\Lambda$--dependence (in the small $\Lambda$ 
limit) as that found in the string model \cite{GV}.  

This concludes our discussion on the 
model containing a cosmological constant in the gravitational 
sector of the theory. 
In the following Section, we will consider a second model that 
contains a pre--big bang phase.  

\section{A Cosmological Constant 
in the Matter Sector}

\setcounter{equation}{0}

\def\theequation{\thesection.\arabic{equation}}

In this Section, we  consider a model with
an effective  cosmological constant in the matter sector 
of the theory. Such a term is present 
when a second, self--interacting scalar 
field becomes trapped in a metastable, false vacuum  state. This energy 
is formally equivalent to a dilaton potential of the form 
$\Lambda =\lambda e^{\Phi}$ in Eq. (\ref{simpleaction}), 
where $\lambda$ is an arbitrary, positive constant. 

The field equations derived from Eq. (\ref{simpleaction}) for this 
form of $\Lambda (\Phi)$ are given by
\bea
\label{fielda}
6 \dot{\alpha}^2 -4 \dot{\alpha}\dot{\Phi} +(2+\omega ) \dot{\Phi}^2 
+4 \ddot{\alpha} -2\ddot{\Phi} =2 \lambda e^{\Phi} \\
\label{fieldb}
12 \dot{\alpha}^2 +6 \omega \dot{\alpha}\dot{\Phi}  -\omega \dot{\Phi}^2 
+6 \ddot{\alpha} +2 \omega \ddot {\Phi} =0 \\
\label{fieldc}
6 \dot{\alpha}^2 -6 \dot{\alpha}\dot{\Phi} -\omega \dot{\Phi}^2  = 
2 \lambda e^{\Phi}
\eea
Subtracting Eq. (\ref{fieldc}) from  (\ref{fielda}) 
eliminates any direct dependence on the cosmological constant:
\be
\label{fieldd}
2\ddot{\alpha} -\ddot{\Phi} +\dot{\alpha}\dot{\Phi} +(1+\omega ) \dot{\Phi}^2
=0
\ee
Subtracting  Eq. (\ref{fieldb}) from (\ref{fieldd}) then implies that 
\be
\label{fielde}
4 \ddot{\alpha} +(1+2 \omega ) \ddot{\Phi} +12 \dot{\alpha}^2 +(6 \omega 
-1) \dot{\alpha}\dot{\Phi} -(1+2\omega ) \dot{\Phi}^2 =0
\ee

Eq. (\ref{fielde}) admits the first integral
\be
\label{integrala}
e^{3\alpha -\Phi}  \left[ 4\dot{\alpha} +(1+2\omega )  \dot{\Phi} 
\right] = \gamma
\ee
where $\gamma$ is an arbitrary constant. 
This constraint is very useful, because it implies that a 
new variable $\chi$, proportional to the term in the 
square brackets, may be introduced. The constant $\gamma$ may then be 
interpreted as the momentum conjugate to this variable.  Consequently, 
the Lagrangian will be 
independent of $\chi$, since $\gamma$ 
is time--independent. Thus, the effective potential in the corresponding 
Wheeler--DeWitt equation will also be 
independent of $\chi$ and this implies that the wavefunction of the Universe 
will be an eigenstate of the momentum operator associated with $\chi$. 

In view of this,  we define the coordinate pair
\bea
\chi \equiv 4 \alpha +(1+2\omega )\Phi \nonumber \\
\psi \equiv \Phi -6 \alpha
\eea
It also proves convenient to define a new time variable
\be
\tau \equiv \int dt e^{3 \alpha (t)}
\ee
since the Wheeler--DeWitt equation does not depend on the specific 
choice of lapse function. 
It follows that action (\ref{simpleaction}) transforms to 
\be
\label{chiaction}
S=\int d\tau \left[ \frac{e^{-\psi}}{4(5+6 \omega)} \left( 
6 \chi '^2 -2(2\omega +3) \psi '^2 \right) -2 \lambda \right]
\ee
for all $\omega \ne -5/6$, where a prime denotes differentiation 
with respect to $\tau$. The transformation 
is unphysical if $\omega = -5/6$ and the significance of this value 
becomes apparent in the Einstein frame. This model is conformally equivalent 
to Einstein gravity minimally coupled to a scalar field 
with an exponential, self--interaction potential. One can show that the 
functional form of the attractor solution in the Einstein frame 
changes when $\omega =-5/6$ \cite{liddle}. 
For $\omega < -5/6$, the potential is so steep that it rapidly redshifts 
to zero and the field effectively becomes massless. For $\omega >-5/6$,
on the other hand,  
the attractor is given by the well known power law solution, where the 
kinetic and potential energies of the field  redshift at the same rate. 

The field equations (\ref{fielda})--(\ref{fieldc}) have been solved 
by Barrow and Maeda in terms of the parametric time $\tau$ for $\omega > 
-5/6$ \cite{BM}. There exist two distinct solutions given by
\bea
\label{exactsolution}
e^{\alpha} = \tau^{\delta_{\pm}} \left( \tau +\tau_1 \right)^{\delta_{\mp}}
\nonumber \\
e^{-\Phi} = \lambda \left( \frac{5+6\omega}{2\omega +3} \right) 
\tau^{\sigma_{\pm}} \left( \tau +\tau_1 \right)^{\sigma_{\mp}} 
\eea
where
\bea
\label{delta}
\delta_{\pm} \equiv \frac{\omega}{3\left( 1+2\omega \pm \sqrt{1+2\omega /3} 
\right) } \nonumber \\
\sigma_{\pm} \equiv \frac{1\pm \sqrt{1+2\omega /3}}{1+2 \omega 
\pm \sqrt{1+2 \omega /3}}
\eea
and 
\be
\tau_1 \equiv \mp \frac{\gamma}{\lambda}  \left( 1+\frac{2\omega}{3} \right)^{-1/2}
\left( \frac{2\omega +3}{5+6\omega}  \right)
\ee
is assumed to be an arbitrary, semi--positive definite  constant. 
It follows that each branch is characterized by equal and opposite 
values of $\gamma$. As $\tau \rightarrow +\infty$, 
both solutions asymptotically approach the 
attractor solution $e^{\alpha} \propto t^{\omega +1/2}$. 
The attractor is recovered by specifying  $\gamma =0$ in the 
first integral (\ref{integrala}). 
There is a curvature singularity at $\tau =0$ and 
Eq. (\ref{exactsolution}) approaches 
the free field solution (\ref{vacuum}) 
as $\tau \rightarrow 0$, i.e., 
$e^{\alpha} \propto  \tau^{\delta_{\pm}} \propto 
t^{p_{\mp}}$. 
Both branches therefore show  qualitatively different behaviour in the 
vicinity of the curvature singularity. The 
constant $\tau_1$ determines the time interval during which the dynamics 
is dominated by the dilaton's kinetic energy. 

We shall consider the example where $\omega =-1/2$. This is interesting 
from a physical point of view, because it corresponds to a dimensionally 
reduced version of six--dimensional Einstein gravity with a 
two--dimensional, isotropic, 
Ricci--flat internal space. In this model, the dilaton 
field is related to the radius of the internal space $b$ by 
$\Phi =-2 \ln b$ \cite{freund}. Eq. (\ref{integrala}) 
implies that the scale factor exhibits no  turning points
when $\omega =-1/2$. Consequently, 
the two branches in Eq. (\ref{exactsolution}) represent 
cosmologies that either expand or contract indefinitely and 
the qualitative behaviour of the solutions for 
all time can therefore be determined from the evolution of the 
scale factor in the high curvature regime. This is given by
$\alpha \propto p_{\pm} \ln t$, 
where $p_{\pm} = -1/(3\mp \sqrt{24})$,  and the $(p_+)$-- and 
$(p_-)$--branches therefore 
represent expanding and contracting solutions. 
However, the time reversal of the contracting solution 
generates a new, expanding solution that is defined 
over $t<0$. 

This implies that the $(p_-)$-- and $(p_+)$--branches 
may be viewed as pre-- and post--big bang solutions, respectively. 
The qualitative behaviour of the scale 
factor in each branch is similar 
to that of the string model considered in Section 
3, where $\omega =-1$. The scale factor 
tends to a finite constant in the low energy limit $(|t| \rightarrow \infty
)$ and diverges as $t \rightarrow 0^-$, as shown in Figure 2b.
The two branches are given in terms of $\chi$ and $\psi$ by 
\bea
\label{prepost}
e^{\chi^{(\pm )}} = \left( \mp \tau \right)^{\mp \sqrt{2/3}} 
\left( \mp \tau + \tau_1 \right)^{\pm \sqrt{2/3}} \nonumber \\
e^{-\psi^{(\pm )}} = \lambda \left( \mp \tau \right) \left( 
\mp \tau + \tau_1 \right)
\eea
and the momenta conjugate to these variables are 
\be
\label{conjugate}
p_{\chi}^{(\pm )} = (3/2)^{1/2} \lambda \tau_1 , \qquad 
p_{\psi}^{(\pm )} = \lambda \left( 2 \tau \mp \tau_1 \right)
\ee
It follows, therefore, that the 
two branches have equal and opposite momentum with 
respect to the shifted dilaton $\psi$. This momentum 
tends to a constant value, 
$p_{\psi}^{(\pm )} \approx 
\mp \lambda \tau_1$, in the region near to 
the curvature singularity $( |\tau | \rightarrow 0)$ and 
is given by  
\be
\label{pvarphi}
\lim_{\psi \rightarrow -\infty} p_{\psi}^{(\pm )} 
= \mp 2 \sqrt{\lambda} e^{-\psi^{(\pm )}/2}
\ee
in the low energy limit
$( |\tau | \rightarrow \infty )$. 

The quantum analysis for this model 
is similar to that considered in the previous Section. The cosmology 
is quantized by identifying the conjugate momenta with the operators 
$\hat{p}_{\chi} =i\partial /\partial \chi$ 
and $\hat{p}_{\psi} =i\partial /\partial \psi$ in the 
Hamiltonian constraint $H_0= p_{\chi} \chi' + 
p_{\psi} \psi' -L =0$. The Wheeler--DeWitt equation is then given by 
\be
\label{wheeler}
\left[ \frac{\partial^2}{\partial \chi^2} - \frac{3}{2}
\frac{\partial^2}{\partial \psi^2} - 6 \lambda e^{-\psi} 
\right] \Psi =0
\ee
The solution that 
satisfies the tunneling boundary condition 
at the singular boundary of minisuperspace
is given by 
\be
\label{wheelersolution}
\Psi = J_p \left( z \right) e^{- ik \chi}
\ee
where $p \equiv -2i\lambda \tau_1$, $k \equiv \sqrt{3/2} \lambda \tau_1$ 
and $z \equiv 4 \sqrt{\lambda} e^{-\psi /2}$. 
It may be verified that 
the  wavefunction (\ref{wheelersolution}) 
is an eigenfunction of both momentum operators $\hat{p}_{\chi}$ and 
$\hat{p}_{\psi}$ in the high energy limit $(\psi \rightarrow +\infty)$. 
Indeed, the ratio of the eigenvalues is given by 
$p_{\chi}/p_{\psi} = - \sqrt{3/2}$, in agreement with 
the classical momenta (\ref{conjugate}) for the pre--big bang phase. 
Moreover, 
in the low energy limit $(\psi \rightarrow -\infty)$, 
the wavefunction (\ref{wheelersolution}) may 
be expressed as a superposition of two components, 
$\Psi = \Psi^{(+)}_{-\infty} 
+ \Psi^{(-)}_{-\infty}$, where
\be
\Psi^{(\pm )}_{-\infty} = \left( 2 \pi z \right)^{-1/2} 
\exp \left[  -i k \chi \mp i z \pm \frac{i \pi}{4} \pm 
\left( \frac{2}{3} \right)^{1/2} \pi k \right]
\ee
These components are eigenfunctions of $\hat{p}_{\psi}$:
\be
\lim_{\psi \rightarrow -\infty} \hat{p}_{\psi} \Psi^{(\pm )}_{-\infty}
 = \mp  2 \sqrt{\lambda} e^{-\psi /2} \Psi^{(\pm )}_{-\infty}
\ee
and comparison with Eq. (\ref{pvarphi})
implies that they represent the pre-- and post--big bang branches, 
respectively. We may conclude, therefore, 
that there is a non--zero  probability for the quantum transition 
between the two branches. It  is given by 
\be
R=e^{-\sqrt{32/3} \pi k}
\ee

\section{Conclusion}

In this paper we have considered the pre--big 
bang cosmological scenario within the context 
of the Brans--Dicke theory of gravity. 
In the free field model, an 
epoch of superinflationary expansion may occur 
during a pre--big bang phase when $-4/3 < \omega <0$. 
This is an example of pole--like, kinetic inflation, where
the expansion  is driven by the kinetic energy of the dilaton. 

At the quantum cosmological 
level, the two branches are represented by wavefunctions that 
move in opposite spatial directions through the 
configuration space. A transition between the two 
branches may be viewed as a spatial reflection of the wavefunction. 
A non--trivial dilaton potential is required, however, if such a reflection 
is to proceed. 

The pre-- and post--big branches of the free field 
model are related by a generalization 
of the scale factor duality associated with the string effective 
action. This generalization allows us to gain further 
insight into why string theory is related to 
$\omega =-1$ Brans--Dicke theory.  
Scale factor duality in string theory extends the $R$--duality 
of toroidal string compactification \cite{dd}. For example, one can 
consider a closed string propagating on a five--dimensional 
space--time $M_4 \times R^1$, where $M_4$ represents Minkowski space 
and $R^1$ is a flat circle. $R$--duality arises because there is 
no limit to the number of times 
a closed string may wrap itself around the compact fifth dimension. 
This results in a duality in the metric such that 
$\tilde{g}_{55} =g^{-1}_{55}$, where $g_{55}$ is the metric component 
associated with the extra dimension. Eq. (\ref{sfd}) then implies 
that invariance under a direct inversion of the metric components 
is only possible if $\omega =-1$. 

The introduction of a cosmological constant into the gravitational
sector of the theory preserves the scale factor duality symmetry. We 
found classical pre-- and post--big bang solutions that 
are related by this duality but are separated by a curvature singularity. 
We also considered a second model that 
contained an effective 
cosmological constant in the matter sector of the theory. 
In both models, the pre-- and 
post--big bang branches  may be viewed classically as particles moving 
in opposite spatial 
directions through minisuperspace. At the quantum level, 
the wavefunction corresponding to the pre--big bang branch was 
selected by invoking the tunneling boundary condition, as suggested by 
Gasperini {\em et al.} \cite{GMV}. This leads 
to a non--zero probability for a spatial reflection of the 
wavefunction to occur. An approximate calculation  
implied that the probability becomes  higher for a larger cosmological 
constant. In a sense, the reflection 
coefficient represents the probability for the 
birth of our Universe (corresponding to the post--big 
bang branch) out of the pre--big bang phase. 

We have not addressed the question of how the semi--classical 
limit is recovered in this scenario. Although the wavefunction reduces 
to a superposition of two semi--classical wavefunctions in the low--energy 
limit, it represents a superposition of two 
macroscopically different states and this 
does not correspond to classical behaviour. It 
is possible that this question could be resolved by studying the 
decoherence of the wavefunction \cite{decoherence}. Indeed,  
Lukas and Poppe have recently 
analyzed decoherence in the string model $(\omega =-1, \Lambda =0)$ 
by including the effects of inhomogeneous dilaton 
fluctuations \cite{LP}. 
They have shown that decoherence is possible if certain 
conditions are satisifed and have suggested that decoherence itself 
may result in a branch change. 

It should be noted that the 
range of $\omega$ that allows a superinflationary, pre--big 
bang phase is inconsistent with primordial nucleosynthesis 
constraints \cite{SA} and 
solar system tests \cite{solar}. 
These lead to the lower limit of $\omega >500$ at the present epoch. 
Some  modification to the class of models that we have considered 
in this work is therefore required. 
One possible extension would be to consider more 
general scalar--tensor theories of gravity, where $\omega =\omega (\Phi)$ 
is assumed to be some function of the dilaton. Alternatively, 
more complicated dilaton potentials that contain at least 
one global minimum could be invoked. 

Finally, the generation of primordial scalar and tensor perturbations 
should be considered. It has been suggested that the string  
scenario may produce a significant quantity of gravitational 
waves with frequencies accessible to the next generation of 
gravitational wave detectors \cite{stringgw}. This is  
in contrast to the standard, slow--roll inflationary scenario 
\cite{battye}. It would be of interest 
to investigate whether similar conclusions apply for the 
extended scenario considered in this paper. 
The inhomogeneous modes of the dilaton would need to 
be incorporated into the analysis, however, before definitive 
conclusions could be drawn.

\vspace{.7in}

\centerline{\bf Acknowledgments}

\vspace{.1in}

The author is supported by the Particle Physics and Astronomy 
Research Council (PPARC), UK. It is a pleasure to thank Maurizio Gasperini
for helpful communications.

\newpage

\centerline{\bf FIGURE CAPTIONS}

\vspace{.2in}

{\em Figure 1}: The pre-- and post--big bang solutions 
of  the free field model for $-4/3 < \omega < 0$. 

\vspace{.2in}

{\em Figure 2}: The pre-- and post--big bang solutions when 
a cosmological constant is introduced into the gravitational sector of the 
theory. (a) $-1 < \omega <0$. (b) $\omega =-1$. (c) $ -4/3 < \omega 
< -1$. 

\vspace{.2in}

{\em Figure 3}: The evolution of the effective gravitational 
coupling $G_{\rm eff} \propto e^{\Phi}$ when $\Lambda (\Phi)$ is a 
constant and $-4/3 < \omega < 0$. 

\end{document}